\documentclass{nature}
\usepackage{color} 
\usepackage{footnote}
\makesavenoteenv{tabular}
\makesavenoteenv{table}
\usepackage{amsfonts}

\usepackage{graphicx}
\makeatletter
\let\saved@includegraphics\includegraphics
\AtBeginDocument{\let\includegraphics\saved@includegraphics}
\renewenvironment*{figure}{\@float{figure}}{\end@float}
\makeatother

\title{Segmentation of laterally symmetric overlapping  objects: application to images of collective animal behavior}

\author{Kirill Lonhus, Dalibor \v{S}tys, Mohammadmehdi Saberioon \& Renata Rycht\'{a}rikov\'{a}}

\begin{document}
\maketitle

\begin{affiliations}
 \item University of South Bohemia in \v{C}esk\'{e} Bud\v{e}jovice, Faculty of Fisheries and Protection of Waters, South Bohemian Research Center of Aquaculture and Biodiversity of Hydrocenoses, Kompetenzzentrum MechanoBiologie in Regenerativer Medizin, Institute of Complex Systems, Z\'{a}mek 136, 373 33 Nov\'{e} Hrady, Czech Republic.
\end{affiliations}

\begin{abstract}
Video analysis is currently the main non-intrusive method for the study of collective behavior. However, 3D-to-2D projection leads to overlapping of observed objects. The situation is further complicated by the absence of stall shapes for the majority of living objects. Fortunately, living objects often possess a certain symmetry which was used as a basis for morphological fingerprinting. This technique allowed us to record forms of symmetrical objects in a pose-invariant way. When combined with image skeletonization, this gives a robust, nonlinear, optimization-free, and fast method for detection of overlapping objects, even without any rigid pattern. This novel method was verified on fish (European bass, \textit{Dicentrarchus labrax}, and tiger barbs, \textit{Puntius tetrazona}) swimming in a reasonably small tank, which forced them to exhibit a large variety of shapes. Compared with manual detection, the correct number of objects was determined for up to almost $90 \%$ of overlaps, and the mean Dice-S{\o}rensen coefficient was around $0.83$. This implies that this method is feasible in real-life applications such as toxicity testing.
\end{abstract}


\begin{figure}
\end{figure}

\section{INTRODUCTION}




The study of collective behavior is a challenging task for any method of individual object tracking. Since the artificial marking of individuals, e.g., by an electronic device, can affect their behavior\cite{Dennis_2008}, it is necessary to use nonintrusive methods. Due to its excellent spatial and time resolution, video analysis is the most prominent among these methods. However, the disadvantage of this method is that it only creates a 2D projection of observed objects that are, in reality, always 3D. Indeed, the 2D projection of the whole space unavoidably leads to overlapping of objects in the image. Moreover, information about the shapes and textures of objects is irrecoverably lost. This can be crucial in tracking of individual objects, especially if the density of the objects is high\cite{Delcourt2012}. This holds mainly for video analysis of typical collective behavior like fish schooling, bird flocking, or crowds of people, when the individuals are overlapping most of the time\cite{Li2015, Kok2016, Jolles2017}. 

The most commonly used solutions of the problem of multiple object tracking using machine vision systems still ignore the overlaps. Nevertheless, the overlaps can be typically detected from morphological parameters (e.g., area, perimeter, eccentricity, and their combinations) of the image binary mask, and the trajectories of the individuals are then reconstructed using track extrapolation\cite{Delcourt_2009}, particle filter\cite{Morais2005}, or, more frequently, Kalman movement prediction\cite{Raj2016, Tang2013}. These methods work well for less crowded scenes where, in addition, the mobility of the individuals is low. Solutions that are directly aimed at observing the interactions in groups usually utilize texture matching before and after collision\cite{Perez-Escudero2014} instead of movement prediction. This introduces some robustness into the approach but important information about the movement is not utilized.

None of these methods mentioned above can track individuals in very dense scenes, where the objects are overlapping most of the time. For these complicated cases, the method of rigid pattern matching can be used\cite{Terayama_2016}. This method requires the shape of the object to be invariant, which is rarely the case in practical applications, and thus the detection rate of this method is low. Most of the more advanced solutions include complex nonlinear optimizations to fit a set of ellipsoids around a fish's central line in 3D\cite{Butail2011}. This approach works for dense scenes but the model used contains many degrees of freedom, which leads to significant inaccuracies, and requires high-resolution imaging and a high frame rate. In addition, this method is not single-image and requires knowledge of the number of objects and the previous state of the system. But the previous state of the system can unavoidably lead to the error propagation and the number of objects is not always known, for example, in cases where the observed volume contains hideouts or is not fully open. A method which is reliable under such conditions is crucial for the study of collective behavior, since many interesting behavioral patterns are observed mainly in complex environments with obstacles, hideouts, and inanimate models\cite{Saverino2008, Calfee2016}.

In this paper, for the very first time, we propose a robust, nonlinear, optimization-free, and pose-invariant way of solving laterally symmetric, overlapping objects. The method is truly single-image and---as verified on image sets of fish schools---tolerant of severe data corruption and low image resolution.

\section{BINARY IMAGE DATA ACQUISITION}

{To describe and validate the method, two species of fish were video-observed and analyzed: Relatively large European bass (\textit{Dicentrarchus labrax}); and tiger barb (\textit{Puntius tetrazona}), a popular aquarium~fish.}

\subsection{\textbf{Design of Experiments on Fish}} 

Two experiments were conducted with European bass. In the first experiment, only one individual was recorded in a relatively small ($d$ = 3 m) circular tank by an IR camera with $1280\times 1024$ resolution. The scene was illuminated by an 830 nm, 60 mW IR diode placed above the tank. The required set of individual object fingerprints was collected during this experiment. In the second experiment, 20~individuals of European bass swam together in the same tank. We note that the usage of an IR camera is not essential to the method, as the same results may be obtained with an ordinary video camera. The contours of the detected individuals, as well as of the overlaps, are very noisy due to the dependency of the fish texture coloration in the IR at the depth in which the fish are swimming. The average length and width of the fish projected on the camera was 190 px and 110 px (3 $\mu$m$^{2}$/px), respectively, in 12-bit~depth.

For tiger barb, only one experiment was conducted, with 6 fish individuals swimming in a small (375 $\times$ 210~mm$^2$) tank. All sides of the aquarium were observed simultaneously using a mirror system. Data collection was done separately for each view of the aquarium. The typical projected size of the fish in the bottom and top views of the tank was $50\times 30$ px (5 $\mu$m$^{2}$/px), in 12-bpc color depth.

All experimental manipulations were conducted according to the principles of the Institutional Animal Care and Use Committee (IACUC) of the University of South Bohemia in \v{C}esk\'{e} Bud\v{e}jovice, Faculty of Fisheries and Protection of Waters, Vodňany, Czech Republic, based on the EU harmonized animal welfare act of the Czech Republic. The described research was reviewed and approved by the Ministry of Education, Youth and Sports Committee (MSMT--8792/2017--2). {The original color and grayscale datasets are available upon request from the authors.}

\subsection{\textbf{Acquisition of Binary Masks of Fish in Overlap}}

For both datasets, foreground detection was performed using the Gaussian Mixture Model\cite{Stauffera, KaewTraKulPong2002} with 8 Gaussians. Two consecutive image dilations\cite{Boomgaard1992} with a {1~px diamond structural element fixed the detection artifacts.}

{After that, automatic classification was used for distinguishing binary masks of individual fish from binary masks of overlapping fish. In the case of European bass}, it was assumed that the analyzed object was a single fish if the area of the produced mask obeyed:
\begin{equation}
A\subset \mathbb{E}(A_s) \pm \frac{3}{2}std(A_b),
\end{equation}
where $A$ is the area of mask, $A_b$ are the areas of all detected blobs, $std$ is the standard deviation, and {$\mathbb{E}$ is the expected value}. If the mask area fell outside this range, the object was classed as overlapping~individuals.

For tiger barb we used a more advanced classification method. We assume that a bivariate distribution of area and perimeters of the binary masks is a mixture of two Gaussian mixtures. The~first component (peak) of the mixture corresponds to the images of fish individuals, and has a lower mean value and standard deviation, whereas the second component can be related to the overlapping. After fitting, a posterior probability for each of the images was calculated. We treated the binary image as a fish individual or a fish overlap if the corresponding probability was greater than 0.5.

The binary masks with the overlapped fish were used as the input data for the proposed algorithm~(Figure~1), described in detail in the following section.

\begin{figure}
\centering
\includegraphics[width=0.7\textwidth]{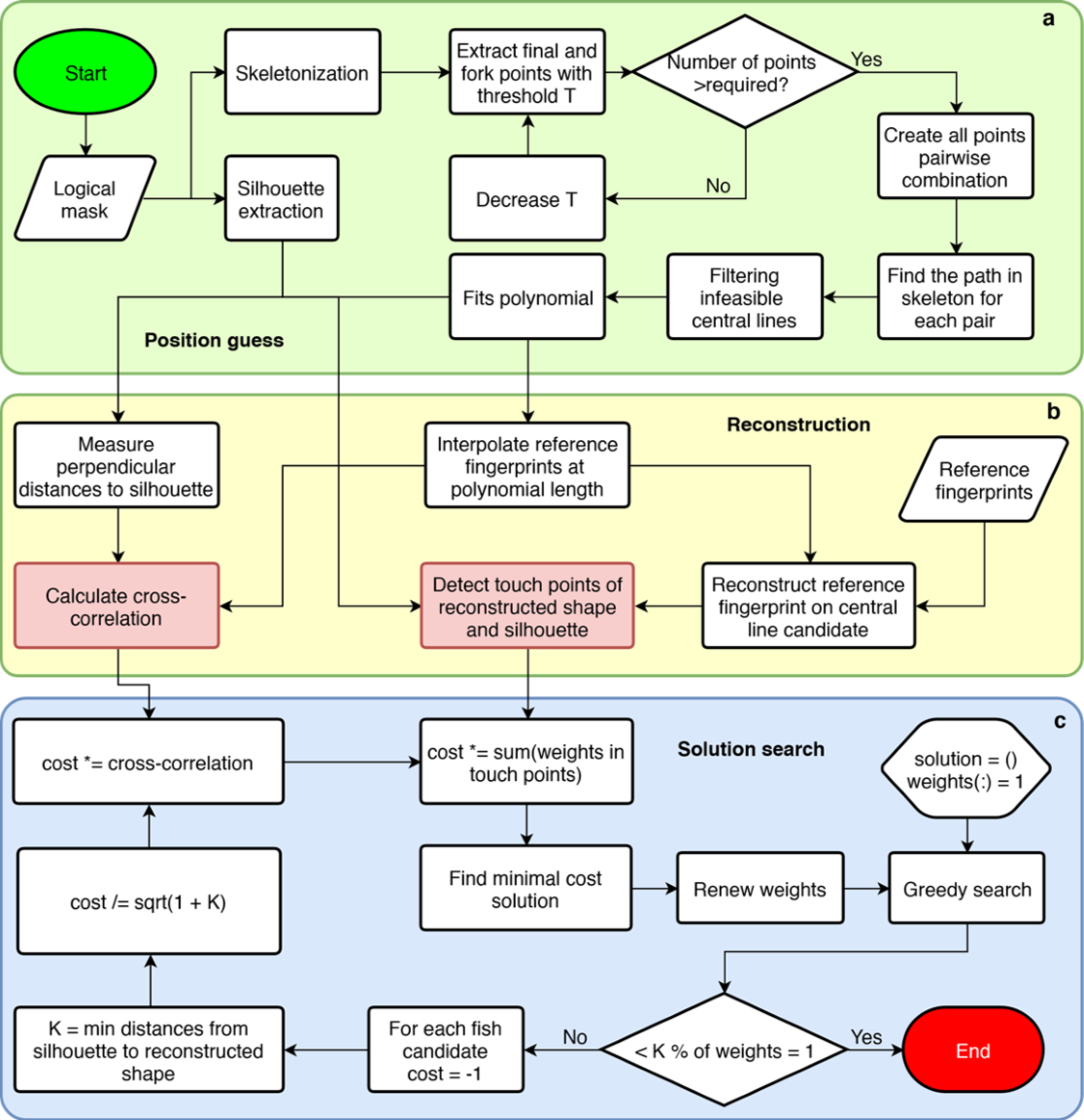}
\caption{Algorithm for segmentation of fish overlaps. (\textbf{a}) Phase of the position guess which includes choosing of the desired level of details, construction of all possible central lines, and filtering of unsuitable central lines out. (\textbf{b}) Fingerprint reconstruction followed by measurement of the fitting accuracy. (\textbf{c}) Greedy search for finding the best solution among the precalculated ones.}
\label{Fig1}
\end{figure}

\section{SEGMENTATION OF BINARY IMAGES OF SYMMETRIC OBJECTS IN OVERLAP} \label{main_algorithm}
\vspace{-6pt}
\subsection{\textbf{Morphological Fingerprinting}} \label{fingerprinting}

Most fish $species$ (and other vertebrates) are bilaterally symmetric\cite{Hollo2015}, i.e., only the transverse plane divides them into two asymmetrical parts. This allows one to describe the morphology of a fish according to the distances of the boundaries of its body from its central line. To proceed with this approach, all binary masks must first adopt the same orientation. There are two degrees of freedom in the orientation of the fish image---rotation around the center and reflection. The angle of the fish binary mask is defined as the angle between the ellipsoidal fit of its major axis and the $ox$ axis. To ensure the same orientation of all images, the algorithm rotates the image to the zero angle. This results in horizontal orientation of the fish and implies two cases where the fish head is located in either the right or left half of the image. Flipping the images in order to locate the center of mass of the binary mask in the left part of the image completes the standardization of the images, so that the masks all have the same orientation. 

After this 'normalization', the positions of central lines are determined from the distance transform. This technique converts the binary silhouette of the normalized fish image to a matrix of the minimal distances from corresponding pixels to the contour of the mask. To do this, an algorithm with time linear complexity\cite{Maurer2003} was used. The central line was defined for the whole fish silhouette in such a way that the $X$ coordinates correspond to the orders of the columns in the distance matrix, and the $Y$ coordinates are defined as the indices of {the rows with the minima in the relevant columns}. The~obtained line was fitted by a low-rank (2--4) polynomial and corresponds to the intuitive estimation of the fish line of symmetry. Its length reflects the true fish length, independent of its pose. To determine the shape, the polynomial fitting of the central line was divided into $N_e$ equidistant points (50 points in our datasets). In order to reduce the computational complexity, piecewise approximation in these equidistant points substituted solving of the integral equations. The~perpendicular distances from these points to the contour define the shape of the fish (Figure~2a,f). As with the length of the central line, these distances are pose-invariant and reflect the morphological properties of the fish. The~average fingerprint can be easily calculated from the relevant mean values. Moreover, this approach to fingerprinting is reversible, and the silhouette of the fish can be constructed for any, even previously unseen, poses (Figure~2e).

The set of pose-invariant fingerprints is an alternative to the rigid pattern used in other models, and reflects the strong dependence of fish proportions on the length of the fish central line. From the biological point of view, this length corresponds to the fish size\cite{MojekwuTOAnumudu2015}.

\begin{figure}
\centering
\includegraphics[width=0.95\textwidth]{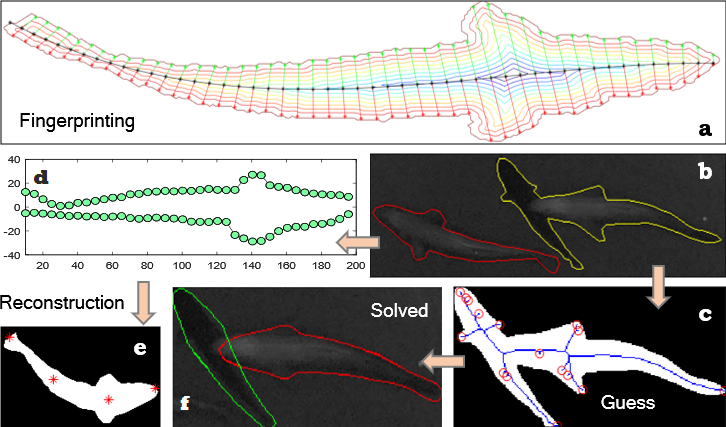}
\caption{The procedure for the segmentation of fish (European bass) overlaps. (\textbf{a}) The process of obtaining a fish fingerprint from contours. (\textbf{b}) The original image of fish overlapping. (\textbf{c}) The skeletonization with desired level of detail of a fish overlap, segmented from image (\textbf{b}). (\textbf{d}) The fish fingerprint. (\textbf{e}) Reconstruction of the fish shape from an arbitrary pose. (\textbf{f}) The final solution.}
\label{Fig2}
\end{figure}

\subsection{\textbf{Dynamic Pattern Extraction}}

The reconstruction of the fish shape using a line of symmetry requires preliminary image treatments. The foreground detection used to obtain the binary mask for fingerprinting can give plausible but not reliable results. These irrelevant fingerprints ($1~\%$ of the lowest and highest lines) were eliminated from the dataset. The resultant dependency of the shape on the length was smoothed using a robust $k$-dimensional spline\cite{Garcia2010}. The obtained data was stored in the form of an $F\times P$ matrix, where $F$ is the number of remaining lengths and $P$ is the count of the points per central line. Together with this matrix, the values of lengths are stored in a column vector. 

For a given length of the central line, this form of data storage allows fast interpolation of the fish shape. In order to eliminate undesirable oscillations, which may produce negative distances, a linear interpolation was used.

\subsection{\textbf{Position Guess}}

Searching for fish individuals in overlap is a challenging task, and implies a complex nonlinear optimization which can be avoided if a set of possible solutions is known (Figure~1a,b). To obtain these solutions, we assume that the fish central lines coincide with the fish image skeletons. The positions and strength of edges are sensitive to the smoothness of the skeleton. Due to this fact, we applied a balanced skeletonization\cite{Eede2006}. To ensure the same level of detail (LoD) for different overlaps, two kinds of skeleton special points---fork and end---were defined. The LoD denotes a number of these points. The subroutine of the equalization process reduced the skeletonization threshold by a fixed step until the LoD exceeded the desired LoD. The optimal count of skeleton points is dependent on the image size. We incorporate it into the heuristic formula $T=8+0.5\sqrt{min(\mathcal{W}, \mathcal{H})}$, where $\mathcal{W}$ is the image width and $\mathcal{H}$ is the image height.

Not all variants of the central lines are relevant for the computation, and so overly short or overly long central lines were removed. The range of acceptable lengths was defined as \begin{equation}
\label{eq:lext}
L_e = Mdn(lines)\pm U\cdot std(lines),
\end{equation}
where $lines$ are the lengths of the central lines in the training set and $U = 1.5$ is related to the dispersion of fish sizes in the experiment. The central line is assumed to be polynomial with coordinates of a high uniqueness ($>$0.8) along at least one dimension, i.e., 80\% of points form the polynomial (nonconstant)~function. 

The variants of the central line that fulfilled the above-mentioned criterion underwent polynomial fitting of order 4. Whereas the nearly vertical lines could not be fitted well by a polynomial, their transposed representations could. To determine the correct orientation, the original and the transposed data were fitted by polynomial and evaluated for the minimum of the standard deviation in the fitting~discrepancy.

\subsection{\textbf{Solution Search}} \label{solution}

The most accurate fits were found when we applied a simple greedy search to the known set of solutions (Figure~1c). However, first, two variables had to be calculated: The unknown count of objects and the cost function of the greedy search. The unknown count of objects was resolved by introducing a stop criterion. As can be understood intuitively, this criterion is related to the optimal coverage of overlaps by the reconstructed fingerprints. The introduction of this criterion requires the introduction of weights for overlapping pixels. The weights can vary from 1 (absence of a reconstructed object in the vicinity of the corresponding contour pixel) to 0 (ideal coincidence). The stop criterion was defined~as
\begin{equation}
isEnough = \frac{1}{N}\sum_{n=1}^{N}[W_n = 1] < \mathcal{R},
\end{equation}
where $N$ is the count of pixels in an overlapping contour, $W_n$ is the weight of the $n$-th pixel in the contour, and $\mathcal{R}$ is the robustness of the method. We used $\mathcal{R} = 0.22$, which means the algorithm will stop when at least $78\%$ of the contour points are covered. Higher or lower values of this parameter provide more false negative or positive results, respectively.

To evaluate the accuracy of the selected central lines, the shape of the object had to be reconstructed from each central line. The procedure is inverse to the fingerprint acquisition but starts similarly, by dividing the polynomial representation of the central line into  $N_e$ equidistant (along the curve) points. The reconstructed shape is composed of points located within a certain perpendicular distance from the node points along the central line (Figure~2e). The distances are defined by the (interpolated) reference fingerprint, with the same length of the central line as the selected solution. The data for the interpolation was taken from the initial fingerprint set.

The cost function reflects two main measures of the optimization process: How accurately the reconstructed object fits to the contour and how accurately it covers the overlap. Let us denote the first class of cost functions as local costs; and the latter class as global costs. The global cost is defined as a median of distances between intact points of weights equal to 1 in an overlapping contour and the reconstructed fingerprint contour:
\begin{equation}
\label{eq3}
global = Mdn\left[\sum_{n=1}^{N}\sum_{m=1}^{M} [W_n = 1]\cdot (|D_m - C_n| )\right],
\end{equation}
where $N$ is the number of pixels in the overlapping contour, $M$ is the doubled (below and above) number of equidistant points, $W_n$ is the weight for the $n$-th pixel of the contour, $D_m$ denotes the distance from the central line to the $m$-th point of the reconstructed object contour, and $C_n$ the distance to the $n$-th point of the overlapping contour. The main goal in introducing the global cost function is to eliminate solutions that fit fish-like objects that are, in fact, only a small part of the overlaps, rather than the full overlap itself.

The local cost is based on the distances from the central line to the overlapping contour. The measurement procedure bears similarities to the method of fingerprint reconstruction, and includes dividing the central line into $N_e$ equidistant points and measurement of the perpendicular distances to the overlapping contour. These distances were compared with the reference. In making this comparison, two issues must be addressed---ambiguity in fish orientation and excessively large distances (from central lines) at the places where the objects are in overlap.
The uncertainty in the orientation was eliminated by the comparison of the local cost for the reference distances and `flipped' reference distances (about the `y-axis' of the fitted polynomial) when searching for the polynomial fit that minimized the local cost function. To address the problem of excessively large distances, distances are marked as too large if they are more than twice longer than the reference. The excessively large distances are then eliminated from the comparison by replacing them with the reference.
The normalized cross-correlation with zero lag was used as the measure of similarity. To handle the broken distances, the correlation was decreased about the relative (to the number of equidistant points) number of broken distances. The local cost was then defined as
\begin{equation}
\label{eq4}
local = \max_{s}\left[corr(\mathbb{D}, R_s)\right] - \frac{1}{M}\sum_{m=1}^{M}[D_m > 2 R_{m}],
\end{equation}
where $\mathbb{D}$ is a set of distances measured from the central line to the overlapping contour, $R_m$ is the reference distance, $M$ is the number of doubled distances, and $s$ is the forward or backward orientation of the reference distance.

The global and local cost functions are required simultaneously and the resultant cost contains their product. The remaining issue in searching for solutions---the uniqueness of the solutions---can be resolved by introducing a degree of uniqueness. The degree of uniqueness is defined as the median of the weights of points of overlapping contour which are the closest to the solution:
\begin{equation}
\label{eq5}
uniq = Mdn(W_n),~n=arg\left(\min_{m} |D_m - C_n|\right),
\end{equation}
where $arg$ is the index of the value. The measure of the uniqueness is maximized. This intuitively corresponds to the idea that the solution is correct if it is localized near the points where there are no other solutions.

Combining the above, the total cost was defined as
\begin{equation}
cost = -\frac{local\cdot uniq}{\sqrt{1 + global}},
\end{equation} 
where $global$ and $local$ are global and local costs, respectively, as defined in Equations~4 and 5; and $uniq$ is a degree of solution uniqueness, as introduced in Equation~6.

Searching for solutions includes three main steps: Finding the solution of the minimal cost, renewing the weights, and checking the stopping criterion. To determine the weights, discrepancies between the solution and contour are calculated. The discrepancies are defined as the mean of the distances between the points of solution and the nearest 3 points of the overlapping contour. We~denote this parameter as $fuzziness$. All weights where the minimal distance from the corresponding points to the solution contour was less than three $fuzzinesses$ were divided by the term $3\times fuzziness$. This feedback was aimed to eliminate coincident solutions.

\section{RESULTS AND DISCUSSION}

The morphological fingerprinting technique demonstrates great robustness and stability for both testing datasets: High-resolution (in terms of pixels per object) but noisy images of European bass; and low-resolution but smooth tiger barb images. The mean shapes of the fish in the corresponding datasets are shown in Figure~3. The mean fingerprint of a single European bass is presented for 174 images (Figure~3c). As can be easily seen, for European bass, the standard deviation of the distances shows peaks in two high-variability regions, which are the fins and the tail. This splits the European bass image into three regions of stability along the central line. These regions may be used in one of the possible applications of the method---fingerprinting of individuals. The mean fingerprint (1182 images) of tiger barb (Figure~3a) is presented for six fish individuals. Here, the standard deviation is nearly uniform along the central line which may be interpreted as being due to large differences in the shapes of the fish individuals. In addition, the distances are significantly more noisy than for European bass. This can be ascribed to the usage of the lowest possible image resolution that was still suitable for fingerprinting of the fish.

The standard deviation of the curvature of the mean central line is significantly more interesting than the curvature itself. It shows points where the fish does or does not bend. The standard deviation of the curvature for both species (Figure~3b,d) has two extrema. The minimum corresponds to the position of the fish skull and implies the inability of the fish to bend at this point. The maximum corresponds to the fish inflection point, which is virtually impossible to locate, even by visual inspection of the images. As seen in the plots of standard deviations of the curvature (Figure~3b,d), the studied species' bodies are of significantly different absolute sizes and proportions.

A further benefit of the technique is that it enables one to solve the overlapping. In order to evaluate the method accuracy, the results were compared with manually segmented overlaps ({147}~images of European bass and 187 images of tiger barbs). As the main measures of accuracy, we used the correct determination of the amount of overlapping objects and the deviation in the centroids of objects. The secondary measures of quality were the deviation in the object orientation and the similarity of the computer-aided contours to the manually segmented contours. As the measures of the contour similarity, the Dice-S{\o}rensen coefficient\cite{Dice1945}, {the Jaccard coefficient}\cite{Jaccard1912}, {and the boundary F1 score}\cite{Csurka2013} were chosen. {For both series, the best matching between the manual segmentation and automatic method was obtained for the classical Dice-S{\o}rensen coefficient.} The results of similarity in detection of fish individuals between the proposed method and the manual segmentation is summarized in Table~1.

The discussed method shows great potential and, despite the low quality of the foreground detection, found the correct amount of fish in {almost $90~\%$} of cases for tiger barb. The mean error in the determination of centroids  was {$5.9 \%$} (in units of fish length), and the angular discrepancy was {$4.2^{\circ}$}. The~Dice-S{\o}rensen coefficient was {$0.82$}, which we interpret as a good result because the proposed method is not a segmentation method but rather a reconstruction method and deals with lost and irrecoverable information. For the European bass dataset, the correct amount of fish was determined in $83~\%$ of cases, centroids error was $5.8 \%$, the Dice-S{\o}rensen coefficient was $0.83$, and the angular discrepancy was $5.7^{\circ}$. {The boundary F1 score is lower for European bass, which can be related to the different level of detail. In other words, the images of tiger barb are significantly smaller and their contours are more convex. Angle discrepancies demonstrate the same tendency. Moreover, in the case of extremely low-resolution images, the method works well for dorsal and ventral views of the fish (Figure~4). Therefore, let us note that downscaling the images before processing might be a beneficial strategy for potential process~speed-up.}

The mean time of calculation per overlap was approx. 5 s. The code is written as a prototype in MATLAB and is not fully optimized. A substantial amount of time is consumed by the self-overhead of functions, and thus the method can be significantly optimized.

The binary datasets (for fingerprinting the tiger barb, we present another, shorter, dataset than shown in the text), all codes, verification tools, and the GUI for the fingerprinting and collision solving are available in the supplementary materials\cite{dryad}.

\vskip1pc

\begin{table}
\caption{{Similarity in fish detection between the proposed reconstruction method and the manual segmentation of the fish overlaps}.\label{Tab1}}
\centering
\begin{tabular}{l c c c c c c c}
\hline
\bfseries Series & \bfseries Imgs. & \bfseries Mean BF & \bfseries JAC & \bfseries Dice & \bfseries Count [\%] & \bfseries Centroid e. [\%] & \bfseries Orient. [\boldmath{$^{\circ}$}] \\
\hline
T. barb & 187 & $0.91 \pm 0.10$ & $0.71 \pm 0.09$ & $0.82 \pm 0.07$ & 89.30 & $5.97 \pm 3.78$ & $4.16 \pm 7.42$ \\
E. bass & 147 & $0.45 \pm 0.18$ & $0.72 \pm 0.13$ & $0.83 \pm 0.10$ & 84.35 & $5.52 \pm 4.30$ & $5.69 \pm 11.94$\\
\hline
\end{tabular}

\begin{tabular}{@{}c@{}} 
\multicolumn{1}{p{\textwidth}}{\footnotesize Imgs.---number of images with fish overlaps in the series; Mean BF---boundary F1 score (contour matching score); JAC---Jaccard similarity coefficient; Dice---Dice-S{\o}rensen similarity coefficient; Count---amount of images correctly determined using the automatic method (in \% of the whole image series); Centroid e.---mean error in the determination of centroids (in \% of fish length); Orient.---angular discrepancy (in angular~$^{\circ}$).}
\end{tabular}
\end{table}

\begin{figure}
\centering
\includegraphics[width=0.95\textwidth]{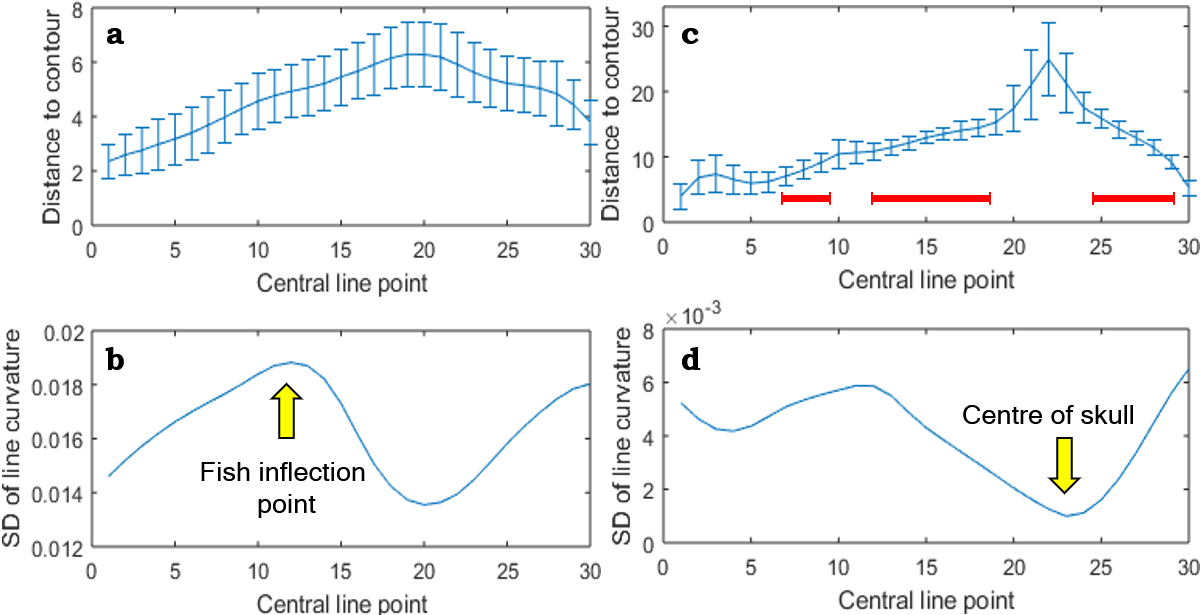}
\caption[]{Mean fingerprints of (\textbf{a}) 6 tiger barbs (for 1182 images) and of (\textbf{c}) a European bass individual (for 174 images). The red lines highlight regions of low variability. (\textbf{b},\textbf{d}) The standard deviations of the central line curvatures for the relevant species.}
\label{Fig3}
\end{figure}

\begin{figure}
\centering
\includegraphics[width=0.95\textwidth]{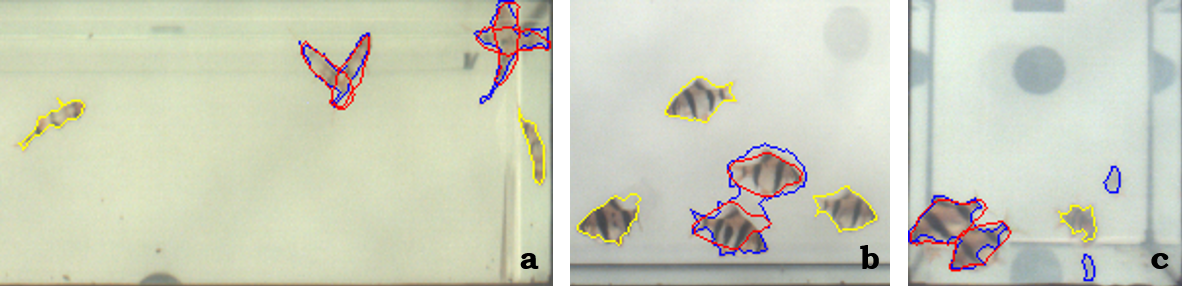}
\caption[]{Solving of overlapping of extremely low-resolution objects. (\textbf{a}) The bottom view of the tank where the fish are strictly symmetrical. (\textbf{b},\textbf{c}) The side views of the tank, where the objects are not symmetrical, but the algorithm can still provide reasonable results.}
\label{Fig4}
\end{figure}

\section{CONCLUSIONS}
{The observation of animal movement using a digital camera is a general approach both in the contemporary ethology research as well as in many practical applications such as observation of fish breeding tanks, cages, or cattle pounds or barns. The presented paper deals with the solution of one of the most complex problems of automated interpretation of such an observation, the division of overlapping animal bodies in a digital image in video-tracking systems. As follows from the literature review, this topic has not been so far sufficiently solved.
 
We present an algorithm providing a robust solution applicable in a wide range of cases found in practice, and show its application in a study of fish school behavior: (1) In a simple observation of relatively large European bass in a recirculating tank by an IR camera (obtained 147 binary images of overlaps); as well as (2) in a very sophisticated ethological setup of an aquarium surrounded by four mirrors and recorded by a fast-speed color camera which provides maximal technically available information on the experiment on tiny tiger barbs (obtained 187 binary images of overlaps). Even in the latter, technically best defined, example, cases of overlapping which are needed to be solved remain. This demonstrate that such an algorithm is unavoidable.}

The solving of overlaps itself is a nontrivial task, especially if objects are nonconvex and have no strict shape. Our method combines only known morphological properties and symmetries of objects with empirical features of image skeletons. Such an approach attains a significant efficiency (hardly attainable by any model-free approach) both in quality and speed. {The segmentation outputs highly correlate with manual/visual object detection (correspondence in up to 90\% cases, similarity indices up to 0.83), mainly in case of relatively small objects of interest as tiger barbs are. Of course, the quality of the proposed segmentation method in general depends on the quality of the binary masks}. The method has been applied to the dorsal and ventral sides of the fish but is supposed to work particularly with semisymmetrical objects such as lateral sides of fish. The same approach with minor changes may be applied (with even greater efficacy than in the case of fish) to other elongated organisms that frequently overlap, e.g., worms and snakes. We believe that the developed method can greatly improve existing systems of tracking and will initiate the development of new ones which will facilitate the processing of previously intractable data.

\section*{{LIST OF SYMBOLS}}

\noindent
\begin{tabular}{@{}ll}
$arg$ & index of the value in computation of the degree of uniqueness\\
$A$ & area of binary mask\\
$A_b$ & area of all detected binary masks, including fish overlaps\\
$cost$ & total cost, i.e., total accuracy of the reconstruction method\\
$C_n$ & distance from the central line to the $n$-th point of the overlapping contour\\
$d$ & diameter of the fish circular tank\\
$\mathbb{D}$ & set of distances from the central line to the reconstructed object contour\\
$D_m$ & distance from the central line to the $m$-th point of the reconstructed object contour\\
$\mathbb{E}$ & expected value\\
$fuziness$ & discrepancy between the solution and the contour\\
$F$ & number of relevant lengths of the fingerprint central line\\
$global$ & global cost, i.e., median of distances between intact points of $W_n=1$ in an overlapping contour\\
& and the reconstructed fingerprint contour\\
$\mathcal{H}$ & image height\\
$isEnough$ & stop criterion in calculation of the unknown number of objects in the solution search\\
$K$ & minimum from the set $\mathbb{D}$ of distances from the central line to the reconstructed object contour\\
$L_e$ & range of acceptable lengths of the central lines in the training set\\
$lines$ & set of lengths of the central lines in the training set\\
$local$ & local cost, i.e., comparison of the perpendicular distances $D_m$ with the overlapping contour with the\\ &  reference length $R_m$ \\
$m$ & order of the doubled equidistant points\\
$M$ & doubled (below and above) number of equidistant points on the fingerprint central line\\
$n$ & order of the pixel in the contour\\
$N$ & count of pixels in an overlapping contour\\
$N_e$ & number of equidistant points on the polynomial representation of the central line\\
$P$ & count of the points per fingerprint central line in dynamic pattern extraction\\
$R_m$ & reference distance in calculation of the local cost\\
$R_s$ & reference distance of the $s$-th, forward or backward, orientation\\
$\mathcal{R}$ & robustness of the solution search method\\
$s$ & orientation, forward or backward, of the reference distance\\
$T$ & optimal count of skeleton points, i.e., the level of detail (LoD)\\
$uniq$ & degree of solution uniqueness of the reconstruction method\\ 
$U$ & dispersion of fish sizes in the experiment\\
$\mathcal{W}$ & image width\\
$W_n$ & weight of the $n$-th pixel in the contour\\
\end{tabular}


\section*{REFERENCES}

\begin{thebibliography}{99}
\bibitem{Dennis_2008}
Dennis, R.~L., Newberry, R.~C., Cheng, H.-W. Estevez, I. Appearance matters: artificial marking alters aggression and stress. \emph{Poultry Sci.} \textbf{87}, 1939--1946 (2008).

\bibitem{Delcourt2012}
Delcourt, J., Denoel, M., Ylieff, M. \& Poncin, P. Video multitracking of fish behaviour: a synthesis and future perspectives. \emph{Fish. Fish.} \textbf{14}, 186--204 (2012).

\bibitem{Li2015}
Li, T. \emph{et~al.} Crowded scene analysis: a survey. \emph{IEEE Trans. Circuits Syst. Video Technol.} \textbf{25}, 367--386 (2015).

\bibitem{Kok2016}
Kok, V.~J., Lim, M.~K. \& Chan, C.~S. Crowd behavior analysis: A review where physics meets biology. \emph{Neurocomputing} \textbf{177}, 342--362 (2016).

\bibitem{Jolles2017}
Jolles, J.~W., Boogert, N.~J., Sridhar, V.~H., Couzin, I.~D. \& Manica, A. Consistent individual differences drive collective behavior and group functioning of schooling fish. \emph{Curr. Biol.} \textbf{27}, 2862--2868.e7 (2017).

\bibitem{Delcourt_2009}
Delcourt, J., Becco, C., Vandewalle, N. \& Poncin, P. A video multitracking system for quantification of individual behavior in a large fish shoal: Advantages and limits. \emph{Beh. Res. Methods} \textbf{41}, 228--235 (2009).

\bibitem{Morais2005}
Morais, E., Campos, M., Padua, F. \& Carceroni, R. Particle filter-based predictive tracking for robust fish counting. In \emph{XVIII Brazilian Symposium on Computer Graphics and Image Processing} (IEEE, 2005).

\bibitem{Raj2016}
Raj, A., Sivaraman, A., Bhowmick, C. \& Verma, N.~K. Object tracking with movement prediction algorithms. In \emph{2016 11th International Conference on Industrial and Information Systems (ICIIS)} (IEEE, 2016).

\bibitem{Tang2013}
qian Tang, W. \& lian Jiang, Y. Target tracking of the robot fish based on adaptive fading Kalman filtering. In \emph{Proceedings 2013 International Conference on Mechatronic Sciences, Electric Engineering and Computer (MEC)} (IEEE, 2013).

\bibitem{Perez-Escudero2014}
Perez-Escudero, A., Vicente-Page, J., Hinz, R.~C., Arganda, S. \& de~Polavieja, G.~G. idTracker: tracking individuals in a group by automatic identification of unmarked animals. \emph{Nat. Methods} \textbf{11}, 743--748 (2014).

\bibitem{Terayama_2016}
Terayama, K., Habe, H. \& aki Sakagami, M.
Multiple fish tracking with an NACA airfoil model for collective behavior analysis.
\emph{IPSJ Trans. Comp. Vis. Appl.} \textbf{8} (2016).

\bibitem{Butail2011}
Butail, S. \& Paley, D.~A. Three-dimensional reconstruction of the fast-start swimming kinematics of densely schooling fish. \emph{J. R. Soc. Interface} \textbf{9}, 77--88 (2011).

\bibitem{Saverino2008}
Saverino, C. \& Gerlai, R. The social zebrafish: Behavioral responses to conspecific, heterospecific, and computer animated fish.\emph{Behav. Brain Res.}\textbf{191}, 77--87 (2008).

\bibitem{Calfee2016}
Calfee, R.~D., Puglis, H.~J., Little, E.~E., Brumbaugh, W.~G. \& Mebane, C.~A. Quantifying fish swimming behavior in response to acute exposure of aqueous copper using computer assisted video and digital image analysis.
\emph{J. Vis. Exp.} (2016).

\bibitem{Stauffera}
Stauffer, C. \& Grimson, W. Adaptive background mixture models for real-time tracking. In \emph{Proceedings 1999 IEEE Computer Society Conference on Computer Vision and Pattern Recognition (Cat. No. PR00149) IEEE Comput. Soc.}).

\bibitem{KaewTraKulPong2002}
KaewTraKulPong, P. \& Bowden, R. An improved adaptive background mixture model for real-time tracking with shadow detection. In \emph{Video-Based Surveillance Systems}, 135--144 (Springer US, 2002).

\bibitem{Boomgaard1992}
van~den Boomgaard, R. \& van Balen, R. Methods for fast morphological image transforms using bitmapped binary images. \emph{CVGIP: Graph. Models Image Process.} \textbf{54}, 252--258 (1992).

\bibitem{Hollo2015}
Holl{\'{o}}, G. A new paradigm for animal symmetry. \emph{Interface Focus} \textbf{5}, 20150032 (2015).

\bibitem{Maurer2003}
Maurer, C., Qi, R. \& Raghavan, V.
A linear time algorithm for computing exact Euclidean distance transforms of binary images in arbitrary dimensions. \emph{IEEE Trans. Pattern Anal. Mach. Intell.} \textbf{25}, 265--270 (2003).

\bibitem{MojekwuTOAnumudu2015}
Mojekwu, T.~O. \& Anumudu, C.~I. Advanced techniques for morphometric analysis in fish. \emph{J. Aquac. Res. Development} \textbf{06}, 354 (2015).

\bibitem{Garcia2010}
Garcia, D. Robust smoothing of gridded data in one and higher dimensions with missing values. \emph{Comput. Stat. Data Anal.} \textbf{54}, 1167--1178 (2010).

\bibitem{Eede2006}
van Eede, M., Macrini, D., Telea, A., Sminchisescu, C. \& Dickinson, S. Canonical skeletons for shape matching. In \emph{18th International Conference on Pattern Recognition} (IEEE, 2006).

\bibitem{Dice1945}
Dice, L.~R. Measures of the amount of ecologic association between species. \emph{Ecology} \textbf{26}, 297--302 (1945).

\bibitem{Jaccard1912}
Jaccard, P. The distribution of the flora in the alpine zone. \emph{New Phytol.}, \textbf{11}, 37--50 (1912).

\bibitem{Csurka2013}
Csurka, G., Larlus, D., \& Perronnin, F. What is a good evaluation measure for semantic segmentation? In \emph{Proceedings of the British Machine Vision Conference} 32.1--32.11 (BMVA Press 2013).

\bibitem{dryad}
Lonhus, K., \v{S}tys, D., Saberioon, M. \& Rycht\'{a}rikov\'{a}, R. Segmentation of laterally symmetric overlapping objects: application to images of collective animal behaviour. \emph{Dryad Digit. Repos.} doi:10.5061/dryad.1j29991.

\end{thebibliography}


\begin{addendum}
\item[ACKNOWLEDGEMENTS] This work was supported by the Ministry of Education, Youth and Sports of the Czech Republic---projects CENAKVA (LM2018099) and the CENAKVA Centre Development (No.~CZ.\\1.05/2.1.00/19.0380)---and from the European Regional Development Fund in frame of the project Kompetenzzentrum MechanoBiologie (ATCZ133) in the Interreg V-A Austria--Czech Republic programme. The work was further financed by the TA CR Gama PoC 02-22-\v{S}tys sub-project and by the GAJU 013/2019/Z project. The authors would like to thank Luke Shaw for editing and proofreading the manuscript.

\item[AUTHOR CONTRIBUTION]

Conceptualization, K.L.; data curation, M.S. and K.L.; formal analysis, K.L. and R.R.; funding acquisition, D.\v{S}.; investigation, K.L.; methodology, K.L.; resources, D.\v{S}.; software, K.L.; supervision, D.\v{S}. and R.R.; validation, R.R.; writing---original draft, K.L.; writing---review and editing, R.R., M.S., and D.\v{S}.

\item[COMPETING FINANCIAL INTERESTS]

The authors declare that they have no competing financial interests.

\item[Correspondence]

Correspondence and requests for materials should be addressed to K.L.\linebreak (lonhus@frov.jcu.cz).
\end{addendum}


\end{document}